\documentclass[preprint,prd,amsmath,amssymb,amsthm,nofootinbib]{revtex4}
\usepackage[mathscr]{euscript}
\usepackage{bm}% bold math
\usepackage{graphicx}% Include figure files
\usepackage{subfigure}
\usepackage{appendix}
\usepackage{multirow}
\usepackage{floatrow}
\usepackage{array}
\usepackage[colorlinks=true,linkcolor=red,urlcolor=blue]{hyperref}
\newcommand{\bea}{\begin{eqnarray}}
\newcommand{\eea}{\end{eqnarray}}
\newcommand{\beq}{\begin{equation}}
\newcommand{\eeq}{\end{equation}}

\def\/{\over}

\begin{document}
\title{Resonance instability of primordial gravitational waves during inflation in Chern-Simons gravity}
\author{Chengjie Fu${}^{1}$}
\email{fucj@itp.ac.cn}

\author{Jing Liu${}^{2,3}$}
\email{liujing@ucas.ac.cn}

\author{Tao Zhu${}^{4, 5}$}
\email{zhut05@zjut.edu.cn}

\author{Hongwei Yu${}^{1}$}
\email{Corresponding author: hwyu@hunnu.edu.cn}

\author{Puxun Wu${}^{1}$}
\email{Corresponding author: pxwu@hunnu.edu.cn}

\affiliation{
${}^{1}$Department of Physics and Synergetic Innovation Center for Quantum Effects and Applications, Hunan Normal University, Changsha, Hunan 410081, China \\
${}^{2}$School of Fundamental Physics and Mathematical Sciences, Hangzhou Institute for Advanced Study, University of Chinese Academy of Sciences, Hangzhou 310024, China\\
${}^{3}$School of Physical Sciences, University of Chinese Academy of Sciences, Beijing 100049, China \\
${}^{4}$Institute for Theoretical physics \& Cosmology, Zhejiang University of Technology, Hangzhou, Zhejiang 310032, China\\
${}^{5}$United Center for Gravitational Wave Physics (UCGWP), Zhejiang University of Technology, Hangzhou, Zhejiang 310032, China}

%\date{\today}

\begin{abstract}
We investigate axion inflation where the gravitational Chern-Simons term is coupled to a periodic function of the inflaton. We find that tensor perturbations with different polarizations are amplified in different ways by the Chern-Simons coupling. Depending on the model parameters, the resonance amplification results in a  parity-violating peak or a board plateau in the energy spectrum of gravitational waves, and the sharp cutoff in the infrared region constitutes a characteristic distinguishable from stochastic gravitational wave backgrounds produced by matter fields in Einstein gravity.

\end{abstract}

\maketitle

\section{Introduction}
\label{sec1}

Primordial gravitational waves (GWs) from quantum
fluctuations of the tensor modes of the spacetime metric were stretched outside the horizon during inflation, and were then frozen on super-Hubble scales. In Einstein gravity, the amplitude of the power spectrum of primordial tensor perturbations produced in the single-field slow-roll inflationary models
is proportional to the energy density of the Universe~\cite{Liddle:1993ch,Guo:2010mm}, and it can be measurable by CMB polarization experiments since primordial GWs can lead to the B-mode polarization of the cosmic microwave background (CMB) anisotropies.
The Planck 2018 result combined with the BICEP2/Keck Array BK14 data gives  an upper limit on the tensor spectrum which is quantified by the tensor-to-scalar ratio $r_{0.002}<0.064$~\cite{Aghanim:2018eyx} at the CMB scales. Inflationary models with quartic and cubic potentials which predict strong primordial GWs are strongly disfavored by the Planck data~\cite{Akrami:2018odb}. However, at the scales much smaller than the CMB scales, the constraints on tensor perturbations are relatively loose on a board range of frequencies. Different from the GWs sourced by matter fields \cite{Khlebnikov:1997di,Easther:2006gt,Easther:2006vd}, primordial tensor perturbations could be enhanced by Lorentz-violating massive gravity~\cite{Kuroyanagi:2017kfx}, non-attractor phase in generalized G-inflation~\cite{Mylova:2018yap}, quantum gravitational  inflation~\cite{Romania:2011ez} and the thermal history of the early Universe ~\cite{Figueroa:2019paj,Domenech:2020kqm}.

The parameter resonance amplification of scalar perturbations during inflation has been widely discussed~\cite{Cai2018,Cai:2019jah,Cai:2019bmk, Zhu:2018smk,Zhou:2020kkf}. In this paper, we investigate the similar amplification of tensor perturbations with parity-violation caused by the gravitational Chern-Simons term coupled to an axion field that drives inflation while they are deep inside the horizon during inflation. Compactifications in string theory generically predict the existence of axions in 4-dimensional low-energy effective field theory. The gravitational Chern-Simons coupling commonly arises in the string axiverse as the Chern-Pontryagin density~\cite{Jackiw:2003pm,Alexander:2009tp}, which can affect tensor perturbations during their propagation~\cite{Lue:1998mq,Alexander:2004us,Satoh:2007gn,Yoshida:2017cjl,Jung:2020aem,Dyda:2012rj,Soda:2017dsu,Nojiri:2019nar, Qiao:2019hkz}.
Due to the weakness of gravitational interaction, the string corrections are weakly constrained. The energy spectrum of the amplified GWs has a parameter dependent characteristic sharp peak or a board plateau with cutoffs in both the infrared and the ultraviolet regions. In Einstein gravity, a power-law slope appears in the infrared region of the GW energy spectrum sourced by the transverse-traceless component of the matter fields~\cite{Cai:2019cdl,Cai:2017cbj}. Detecting the sharp infrared cutoff of the GW energy density provides an important clue to the Chern-Simons coupling, as well as extra dimensions predicted in string theory. This resonance peak might be observed by ground-based and space-based GW detectors, such as aLIGO \cite{aLIGO}, LISA \cite{LISA}, Taiji \cite{Taiji}, and pulsar timing arrays, such as SKA \cite{SKA}.
It is worthy to mention that, in some parameter space, the string-inspired resonance peak is similar to the noise in the measured sensitivity curve of LIGO~\cite{Martynov:2016fzi,Abbott:2019ebz}, and has the possibility to explain the unknown measured noise which also has a sharp-peak profile. In turn, if the noise is reduced in the aLIGO further observation runs, the absence of such peaks will provide stringent constraints on the Chern-Simons coupling during inflation. Note that the Chern-Simons coupling between an axion field and the gauge fields could trigger the exponential production of helically polarized gauge bosons by the polarization-dependent resonance during the inflation and the post-inflationary preheating \cite{Garretson:1992vt,Prokopec:2001nc,Anber:2009ua,Barnaby:2010vf,Sorbo:2011rz,Adshead:2013qp,Cheng:2015oqa,Adshead:2015pva,Adshead:2016iae,Adshead:2018doq}. These helical gauge bosons can source a polarized spectrum of GWs \cite{Sorbo:2011rz,Adshead:2013qp,Adshead:2018doq}.

The paper is organized as follows. In Sec.~\ref{sec2}, we briefly introduce Chern-Simons gravity and the equation of motion~(EOM) of tensor perturbations.
In Sec.~\ref{sec3}, we study the dynamics of resonant instability of tensor perturbations caused by the Chern-Simons coupling.
In Sec.~\ref{sec4}, we present the numerical results of the amplified tensor perturbations with parity-violation in the axion monodromy inflation model as an example.
Section~\ref{sec5} is devoted to conclusions.

\section{Gravitational waves during inflation in Chern-Simons gravity}
\label{sec2}
In this section, we give a brief introduction to the Chern-Simons gravity, whose action has the following form
\begin{align}\label{action}
\mathcal{S} = \int d^4 x \sqrt{-g}\left[\frac{1}{2\kappa^2}(R+\mathcal{L}_{\rm CS})+\mathcal{L}_{\phi} \right]\;,
\end{align}
where $\kappa^{-1}\equiv M_{\mathrm{p}}=2.4\times10^{18}\;\mathrm{GeV}$ is the reduced Planck mass, $R$ is the Ricci scalar, $\mathcal{L}_{\rm CS}$ is the Lagrangian which contains a Chern-Simons term coupled to a scalar field such
as the universal axion, and $\mathcal{L}_\phi$ is the Lagrangian for the scalar field, which is non-minimally coupled to gravity. As a simple example, we consider the action of the scalar field as
\begin{align}
\mathcal{L}_\phi =  - \frac{1}{2} g^{\mu \nu} \nabla_\mu \phi \nabla_\nu \phi - V(\phi)\;.
\end{align}
Here $V(\phi)$ denotes the potential of the scalar field. The Chern-Simons Lagrangian can be written in the form
\begin{align}
\mathcal{L}_{\rm CS} = \frac{1}{8}\vartheta(\phi) \varepsilon^{\mu\nu\rho\sigma} R_{\rho\sigma \alpha\beta} R^{\alpha \beta}_{\;\;\;\; \mu\nu}\;,
\end{align}
with $\varepsilon_{\rho \sigma \alpha \beta}$ being the Levi-Civit\'{a} tensor defined in terms of the antisymmetric symbol $\epsilon^{\rho \sigma \alpha \beta}$ as $\varepsilon^{\rho \sigma \alpha \beta}=\epsilon^{\rho \sigma \alpha \beta}/\sqrt{-g}$. The Chern-Simons coupling function $\vartheta(\phi)$ should in principle be determined  from the quantum theory of gravity.  In the lack of such a theory at the present, $\vartheta(\phi)$ could be treated as an arbitrary function except for that it should be odd  in $\phi$ as a result of the Green-Schwarz mechanism~\cite{Green}. Since different forms of $\vartheta(\phi)$ affects the propagation of tensor perturbations in different ways, $\vartheta(\phi)$ is expected to  be constrained by the observations of GWs.

In the flat Friedmann-Robertson-Walker Universe, the background metric is given by
\begin{align}
ds^2 = -dt^2 + a(t)^2\delta_{ij}dx^i dx^j\;,
\end{align}
where $a(t)$ denotes the scale factor of the Universe and $t$ represents the cosmic time. The Chern-Simons term in the action (\ref{action}) does not contribute the dynamics of the isotropic and homogeneous Universe, which is dominated by the scalar field $\phi$ that plays the role of the inflaton field during the slow-roll inflation.
In this case, the Friedmann equation, which governs the background evolution, takes exactly the same form as that in General Relativity (GR), i.e.,
\begin{align}
3H^2 = \kappa^2\left[\frac{1}{2}\dot \phi^2 +V(\phi)\right]\;,
\end{align}
where $H$ denotes the Hubble parameter. The evolution of the inflaton field $\phi$ is also the same as that in GR,
\begin{align}
\ddot \phi + 3 H \dot \phi + \frac{dV}{d\phi}=0\;.
\end{align}

Now, let us turn to study the propagation of tensor perturbations on a homogeneous and isotropic background. With tensor perturbations, the spatial metric is written as
\begin{align}
g_{ij}=a(t)^2 (\delta_{ij} + h_{ij}(t, {\bf x}))\;,
\end{align}
where $h_{ij}$ represents the first-order transverse and traceless metric perturbations. Substituting $h_{ij}$ into the action (\ref{action}) and expanding it to the second order, one can derive the action for the tensor perturbations \cite{Qiao:2019wsh},
%In order to derive the equation of motion for tensor perturbations, we substitute $h_{ij}$ into the action (\ref{action}) and expand it to the second order. After tedious calculations, we find
\begin{align}
S^{(2)}_h = \frac{\kappa^2}{2} \int dt d^3 x a^3 \left[ \mathcal{L}_{\rm GR}^{(2)} + \mathcal{L}_{\rm CS}^{(2)}\right],
\end{align}
where
\begin{align}
\mathcal{L}_{\rm GR}^{(2)} &= \frac{1}{4} \left[ \dot h_{ij}^2 - \frac{1}{a^2}(\partial_k h_{ij})^2\right]\;, \\
\mathcal{L}_{\rm CS}^{(2)} &= \frac{1}{4} \left[\frac{\dot{\vartheta}}{a} \epsilon^{ijk}\dot h_{il} \partial_j \dot h_{kl}+ \frac{\dot{\vartheta}}{a^3} \epsilon^{ijk}\partial^2h_{il}  \partial_j h_{kl}\right]\;.
\end{align}
Then taking  the variation  of the above action with respect to $h_{ij}$, one can obtain the equation of motion for $h_{ij}$ \cite{Qiao:2019wsh},
\begin{align}\label{GW_EOM}
\ddot h_{ij} + 3 H \dot h_{ij}  - \frac{1}{a^2} \partial^2 h_{ij}
+ \frac{\epsilon^{ilk}}{a } \partial_l \left[ \dot{\vartheta} \ddot h_{jk} + (2H\dot{\vartheta}+\ddot{\vartheta}) \dot h_{jk} - \frac{\dot{\vartheta}}{a^2} \partial^2 h_{jk} \right]=0\;.
\end{align}
In Chern-Simons gravity, the propagation equations for two circular polarization modes of gravitational waves are decoupled. To study the evolution of $h_{ij}$, we expand it over spatial Fourier harmonics,
\begin{align}
h_{ij}(t,{\bf x}) = \sum_{A={\rm R, L}} \int \frac{d^3 {\bf k}}{(2\pi)^3} h_A(t, {\bf k})e^{i {\bf k}\cdot {\bf x}} e_{ij}^{A}({\bf k})\;,
\end{align}
where $\mathrm{R}$ and $\mathrm{L}$ represent the right-handed and left-handed polarization, respectively. Here $e_{ij}^A$ denote the circular polarization tensors and satisfy the relation
\begin{align}
\epsilon_{i l m} k^l e_{ij}^A = i k\rho_A e^A_{mj}\;,
\end{align}
with $\rho_{\rm R}=1$ and $\rho_{\rm L} =-1$. Then one can write Eq. (\ref{GW_EOM}) in the Fourier space as \cite{Qiao:2019wsh}
\begin{align}\label{GW_EOM_h}
\ddot h_A+ (3+\nu_A)H \dot h_A + \frac{k^2}{a^2} h_A=0\;,
\end{align}
where
\begin{align}
H \nu_A=\frac{\rho_A (k/a) (H \dot \vartheta - \ddot \vartheta)}{1- \rho_A (k/a) \dot \vartheta} = \frac{d\ln[1- \rho_A (k/a) \dot \vartheta]}{dt}\;,
\end{align}
and quantity $\nu_A$ describes the modification of the friction term of gravitational waves, which induces the amplitude birefringence effect of gravitational waves.

\section{Resonance instability of gravitational waves}
\label{sec3}
For the convenience of the following discussion, we define a new variable $X_A$ which satisfies the following relation
\begin{align}
X_A \equiv a^{3/2} \sqrt{1- \rho_A (k/a) \dot \vartheta} h_A\;,
\end{align}
and then Eq. (\ref{GW_EOM_h}) can be rewritten as
\begin{align}\label{GW_EOM_X}
\ddot X_A + \left(\frac{k^2}{a^2} - \frac{\ddot F_A}{F_A}\right) X_A=0\;,
\end{align}
where
\begin{align}
F_A = a^{3/2} \sqrt{1- \rho_A (k/a) \dot \vartheta}\;.
\end{align}
Assuming $|(k/a)\dot\vartheta|\ll 1$, one can obtain
\begin{align}
F_A \simeq a^{3/2}\left( 1 - \frac{\rho_A}{2}\frac{k}{a} \dot\vartheta \right)\;,
\end{align}
\begin{align}\label{m_eff}
\frac{\ddot F_A}{F_A} \simeq  \frac{9}{4}H^2 +\frac{3}{2}\dot H -\frac{\rho_A}{2}\frac{k}{a} \left(\frac{1}{4}H^2\dot\vartheta+\frac{1}{2}\dot H\dot\vartheta + H\ddot\vartheta +\dddot \vartheta  \right)\;.
\end{align}

In this paper, we consider that $\vartheta(\phi)$ has the periodical functional form, and take the following form as a typical representative
\begin{align}\label{coupling}
\vartheta(\phi) = \frac{1}{M^2} \sin\left( \frac{\phi}{\Lambda} \right)\Theta(\phi_s - \phi)\Theta(\phi-\phi_e)\;,
\end{align}
which is purely phenomenological and is adopted in order to investigate the resonant amplification of GWs during inflation.
Here $M$ and $\Lambda$ that have the dimension of mass characterize the magnitude and the oscillation period of $\vartheta(\phi)$, respectively, and $\Theta$ is the Heaviside theta function. Without loss of generality, $\phi_s$ and $\phi_e$ are free parameters to be constrained by observations which denote the field values at the starting and ending point that the Chern-Simons term works. In the subsequent discussions, any quantity with subscript $s$ means that  its value is obtained when $\phi=\phi_s$, likewise for subscript $e$.
Next, we focus on the period when the inflaton goes through from $\phi_s$ to $\phi_e$ and study the effects of a periodic coupling function on the tensor perturbations. In some parameter space, the dominant term of $\ddot F_A/F_A$ in Eq. (\ref{m_eff}) is the $d^3\vartheta/d\phi^3$ one, and then the equation of motion given in Eq. (\ref{GW_EOM_X}) can be reduced to
\begin{align}
\ddot X_A + \left[\frac{k^2}{a^2} - \frac{\rho_A}{2}\frac{k}{a} \frac{\dot\phi^3}{M^2\Lambda^3}\cos\left( \frac{\phi}{\Lambda} \right) \right] X_A=0\;,
\end{align}
where the periodically changing frequency $\omega_k^2 \simeq \frac{k^2}{a^2} - \frac{\rho_A}{2}\frac{k}{a} \frac{\dot\phi^3}{M^2\Lambda^3}\cos\left( \frac{\phi}{\Lambda} \right)$ may trigger the parametric resonance for certain modes.
Considering the case of $|\phi_e-\phi_s| \ll M_\mathrm{p}$, the evolution of the inflaton and the scale factor can be simply described as $\phi = \phi_s + \dot\phi_s(t-t_s)$ and $a=a_se^{H_s(t-t_s)}$ respectively during the period from $t_s$ to $t_e$. After introducing a new time variable $2z = \phi_s/\Lambda + \dot\phi_s(t-t_s)/\Lambda + \pi(1+\rho_A)/2$, the above equation can be cast in the form of the Mathieu equation
\begin{align}\label{Mathieu}
\frac{d^2X_A}{dz^2} + \left[A_k -2q\cos(2z) \right] X_A=0\;,
\end{align}
where
\begin{align}
A_k = \frac{k^2}{k_s^2a^2}\;, \qquad
q = \frac{2k_s k}{M^2a} \;,
\end{align}
with $k_s=|\dot\phi_s/(2\Lambda)|$. The existence of an exponential instability $X_A \propto \exp(\mu_k z)$, known as the parametric resonance, is an important feature of solutions of Eq. (\ref{Mathieu}).
In the case of $|(k/a)\dot\vartheta|\ll 1$, as $q$ denotes the magnitude of $|(k/a)\dot\vartheta|$ and thus $q\ll 1$, the resonance bands are located in some narrow ranges near $A_k\simeq n^2$ ($n=1,2,...$) and each resonance band has a width of order $\triangle k \sim q^n$. Since the first band ($n=1$) is the widest and most enhanced band, in the following analysis we focus on this instability band, $1-q<A_k<1+q$. Moreover, in the present paper we consider $k_s\gg H$, which means that the modes are deep inside the Hubble horizon when they stay in the resonance band.

For the first instability band with $q\ll 1$, the Floquet exponent $\mu_k$ which describes the rate of exponential growth is given by
\begin{align}\label{rate}
\mu_k(t)= \sqrt{\left(\frac{k_sk}{M^2a(t)}\right)^2-\left(\frac{k}{k_sa(t)}-1\right)^2}\;.
\end{align}
Thus the resonance occurs in a narrow band
\begin{align}\label{eq:cdt}
k_{+}<\frac{k}{a(t)}<k_{-}\;,
\end{align}
where $k_{+}=k_s/(1+k_s^2/M^2)$ and $k_{-}=k_s/(1-k_s^2/M^2)$. Eq.~\eqref{eq:cdt} implies that mode $k$ can stay at the resonant band during a finite time since the condition Eq.~\eqref{eq:cdt} is time dependent. A mode can be redshifted into and then out of the resonance band because of the exponential expantion of the Universe. The time of the mode staying in the resonant band can be calculated approximately to be $T_{in}=2k_{s}^{2}/(M^{2}H)$. Since the Chern-Simons term is nonzero only between $t_{s}$ and $t_{e}$, the modes entering the resonance band during the time region ($t_{s}-t_{e}$) are amplified. Using $t_{I}$ and $t_{F}$ to denote the time when the $k$ mode enters and exits the resonance band, respectively, which means $t_s\leq t_I<t_F\leq t_e$, the amplitude of this mode will be amplified by
\begin{align}\label{eq:R}
R_k(t_I,t_F)\equiv \frac{|X_A(t_F)|}{|X_A(t_I)|} \simeq \exp\left(\int^{t_F}_{t_I}\mu_k(t)k_sdt\right)\;,
\end{align}
after it goes through the resonance band. Defining $\mathcal{A}_k(t)=k/(k_sa(t))$, Eq.~\eqref{eq:R} could be written in a more explicit form
\begin{align}
R_k(&\mathcal{A}_k(t_I),\mathcal{A}_k(t_F)) \simeq \exp\left(-\frac{k_s}{H_s}\int^{\mathcal{A}_k(t_F)}_{\mathcal{A}_k(t_I)}\sqrt{\left(\frac{k_s^2}{M^2}\mathcal{A}_k\right)^2-\left(\mathcal{A}_k-1\right)^2}\frac{d\mathcal{A}_k}{\mathcal{A}_k}\right)\; \nonumber \\
= &\exp\Bigg(-\frac{k_s}{H_s}\Bigg[\sqrt{\left(k_s^2\mathcal{A}_k/M^2\right)^2-\left(\mathcal{A}_k-1\right)^2} -
\arctan\frac{\mathcal{A}_k-1}{\sqrt{\left(k_s^2\mathcal{A}_k/M^2\right)^2-\left(\mathcal{A}_k-1\right)^2}} \nonumber \\
&-\frac{1}{\sqrt{1-k_s^4/M^4}}\arctan\frac{1-(1-k_s^4/M^4)\mathcal{A}_k}{\sqrt{1-k_s^4/M^4}\sqrt{\left(k_s^2\mathcal{A}_k/M^2\right)^2-\left(\mathcal{A}_k-1\right)^2}}\Bigg]\Bigg|^{\mathcal{A}_k=\mathcal{A}_k(t_F)}_{\mathcal{A}_k=\mathcal{A}_k(t_I)} \Bigg)\;.
\end{align}

The amplified modes can be divided into three groups: 1) the modes entering the band before $t_{s}$; 2) the modes entering the band after $t_{s}$ and exiting before $t_{e}$; 3) the modes exiting the band after $t_{e}$. These three classes satisfy the following relation, respectively 
\begin{align}
	k_+a_s< k\leq k_-a_s\;, \qquad k_-a_s< k < k_+a_e\;, \qquad k_+a_e\leq k < k_-a_e\;.
\end{align}
For the second class $t_{F}-t_{I}=T_{in}$, while for the first and the third class $t_{F}-t_{I}<T_{in}$ because the time when these modes satisfy Eq.~\eqref{eq:cdt} is not totally between $t_{s}$ and $t_{e}$.

Then the corresponding $\mathcal{A}_k(t_I)$ and $\mathcal{A}_k(t_F)$ can be calculated as
\begin{align}
\mathcal{A}_k(t_I) = \frac{k}{k_sa_s}\;, \qquad \mathcal{A}_k(t_F) = \frac{k_+}{k_s}
\end{align}
for $k_+a_s< k\leq k_-a_s$,
\begin{align}
\mathcal{A}_k(t_I) = \frac{k_-}{k_s}\;, \qquad \mathcal{A}_k(t_F) = \frac{k_+}{k_s}
\end{align}
for $k_-a_s< k < k_+a_e$, and
\begin{align}
\mathcal{A}_k(t_I) = \frac{k_-}{k_s}\;, \qquad \mathcal{A}_k(t_F) = \frac{k}{k_sa_e}
\end{align}
for $k_+a_e\leq k < k_-a_e$. It is interesting to note that the magnification $R_k$ is independent of $k$ for the second group. 

In  the usual slow-roll inflation without a Chern-Simons coupling, the tensor power spectrum evaluated at the horizon crossing [$k=aH$] is given by
\begin{align}
\mathcal{P}_h \equiv 4 \sum_{A=R,L} \frac{k^3}{2\pi^2}|h_A|^2 \simeq \frac{8}{M_{\mathrm{p}}^2}\left(\frac{H}{2\pi}\right)^2\;.
\end{align} 
For the model considered in present paper, since the amplitude of the modes satisfying $k_+a_s< k < k_{-}a_e$ can be amplified, which is denoted  by a factor of $R_k$,  due to the resonance effect, the resulting tensor power spectrum can be written as
\begin{align}\label{Ph}
\mathcal{P}_h\simeq \frac{8}{M_{\mathrm{p}}^2}\left(\frac{H}{2\pi}\right)^2\times\left\{\begin{array}{lcl}
1,\;\;(k\leq k_+a_s \;\& \;k\geq k_-a_e )\\
R_k\left(\frac{k}{k_sa_s},\frac{k_+}{k_s}\right)^2,\;\;(k_+a_s< k\leq k_-a_s)\\
R_k\left(\frac{k_-}{k_s},\frac{k_+}{k_s}\right)^2,\;\;(k_-a_s< k < k_+a_e)\\
R_k\left(\frac{k_-}{k_s},\frac{k}{k_sa_e}\right)^2,\;\;(k_+a_e\leq k < k_-a_e)\\
\end{array}\right. .
\end{align}
At the CMB scales, the amplitude of the tensor power spectrum is limited to less than the order of $10^{-10}$. For the resonance scales, which are smaller than the CMB scales, if tensor perturbations can be enhanced by a certain order of magnitude, the predicted GWs may be probed by the future GW experiments. In the next section, we will study a concrete example by numerical method.

\begin{table}
\caption{The parameter sets for producing the GW signals which could be testable in the future GW experiments.}
\begin{tabular}{>{\centering}p{1.cm}>{\centering}p{2cm}>{\centering}p{2cm}>{\centering}p{1.5cm}>{\centering}p{1.5cm}}
\hline
\hline
$\textit{Case}$ & $\lambda M_{\mathrm{p}}^2/M^{2}$ & $\Lambda/M_{\mathrm{p}}$ & $\phi_s/M_{\mathrm{p}}$ & $\phi_e/M_{\mathrm{p}}$  \tabularnewline
\hline

$1$ & $6\times10^{-9}$ & $2\times10^{-5}$ & $6.88$ & $6.84$ \tabularnewline

$2$ & $5\times10^{-9}$ & $2\times10^{-5}$ & $6$ & $5.88$  \tabularnewline

$3$ & $3\times10^{-9}$ & $2\times10^{-5}$ & $5.10$ & $5.08$ \tabularnewline
\hline
\end{tabular}
\label{table1}
\end{table}

\section{Results}
\label{sec4}
In this section, we present the numerical results by considering the axion monodromy inflation \cite{Silverstein:2008sg,Flauger:2014ana}
\begin{align}
V(\phi) = \lambda M_{\mathrm{p}}^{4-p}\phi^{p}\;
\end{align}
with $p=2/3$. We set the \textit{e}-folding number from the time when the pivot scale $k_\ast=0.002\mathrm{Mpc}^{-1}$ exits the horizon to the end of the inflation as $N_\ast=50$, and the resulting scalar spectral index and tensor-to-scalar ratio are compatible with the current observational constraints \cite{Akrami:2018odb}. Moreover, $\lambda$ is fixed at $4.5\times10^{-10}$ by the amplitude of the curvature perturbations.

\begin{figure}
\centering
\includegraphics[width=0.8\textwidth ]{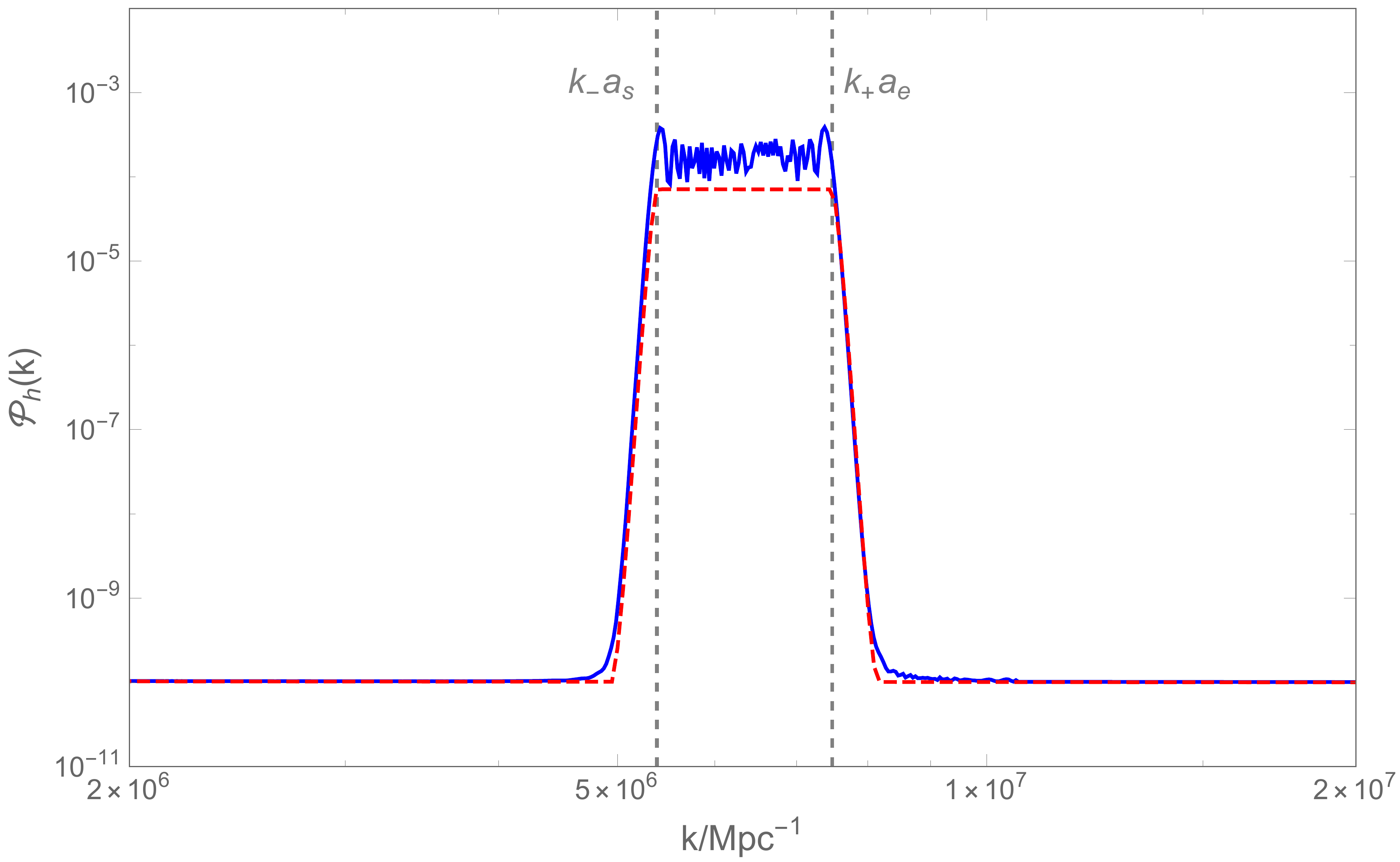}
\caption{\label{fig1} The resulting tensor spectra $\mathcal{P}_h$ for case 1 in Table \ref{table1}. The red dashed and blue solid  lines represent the analytical approximate and exact numerical results, respectively.  }
\end{figure}

Table \ref{table1} shows the three parameter sets we choose for predicting a testable GW background in the future. Taking case $1$ as an example, we plot the tensor spectra $\mathcal{P}_h$ from the analytically approximate result given in Eq. (\ref{Ph}) and the exact numerical result by solving Eq. (\ref{GW_EOM_h}) in Fig. \ref{fig1}. It is easy to see that the numerical result matches the analytical one very well for modes in the $k<k_-a_s$ and $k > k_+a_e$ regimes. For modes within $k_-a_s< k < k_+a_e$, the amplitude of the exact tensor spectrum is larger than that of the approximate one, and it is interesting to observe that the exact tensor spectrum exhibits an irregular structure, instead of the smooth plateau structure predicted by the analytic method. The ripples on the plateau are caused by the phases of $X_{A}$ at the time when $X_{A}$ leaves the resonance band.

\begin{figure}
\centering
\includegraphics[width=0.8\textwidth ]{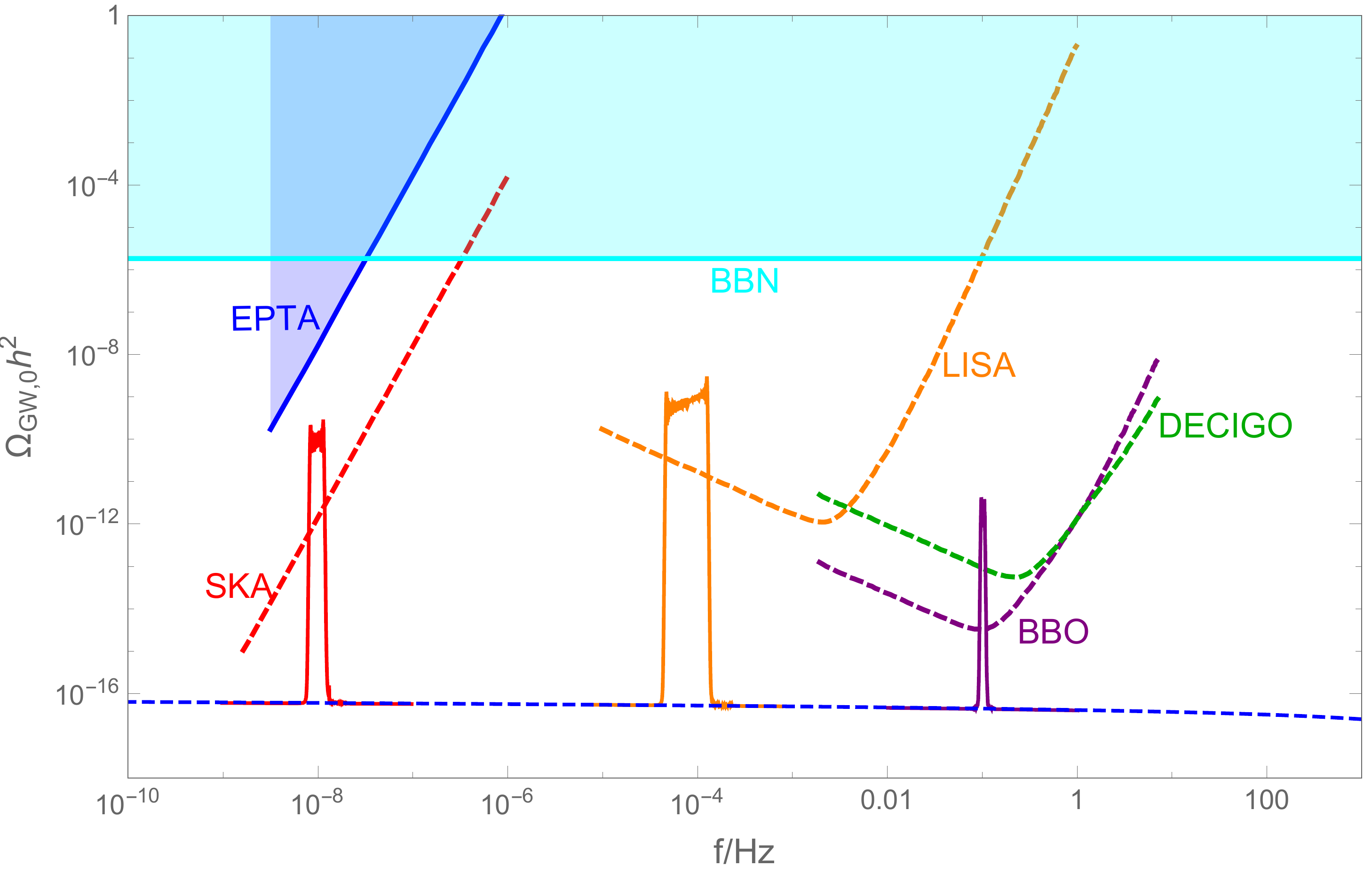}
\caption{\label{fig2} The predicted current energy spectrum of GWs for the axion monodromy inflation model with the gravitational Chern-Simons term. The red/orange/purple solid lines correspond to case 1, 2 and 3 of Table \ref{table1}, respectively. The blue dashed line represent the standard GW energy spectrum predicted by the axion monodromy inflation model. The other dashed lines are the expected sensitivity curve of the future gravitational-wave projects summarized in \cite{Moore12015}. The shaded regions represent the present existing constraints on GWs \cite{Kohri2018,Lentati2015}.}
\end{figure}

For the GWs produced during inflation, the current energy spectrum can be approximately written as \cite{Boyle2008,Zhao2006}
\begin{align}
\Omega_{\mathrm{GW,0}}(k) \simeq 1.08\times10^{-6}\mathcal{P}_h(k)
\end{align}
for the frequency $f>10^{-10}\;\mathrm{Hz}$, and one can relate the current frequency $f$ and the comoving wave number $k$ through
\begin{align}\label{f}
f=1.546\times10^{-15}\frac{k}{1\rm{Mpc}^{-1}}\rm{Hz}\;.
\end{align}
Figure \ref{fig2} shows the current energy spectrum of GWs obtained by numerically solving Eq. (\ref{GW_EOM_h}) for the three cases in Table \ref{table1}. In case 1, the peak of the GW energy spectrum locates in the sensitive region of SKA \cite{SKA}. The peak of the GW energy spectrum for case 2 is above the sensitivity curve of LISA \cite{LISA}. For case 3, the predicted GW spectrum could be simultaneously detected by the deci-hertz interferometer GW observatory (DECIGO) \cite{DECIGO} and the big bang observer (BBO) \cite{BBO}.

\begin{figure}
\centering
\includegraphics[width=0.8\textwidth ]{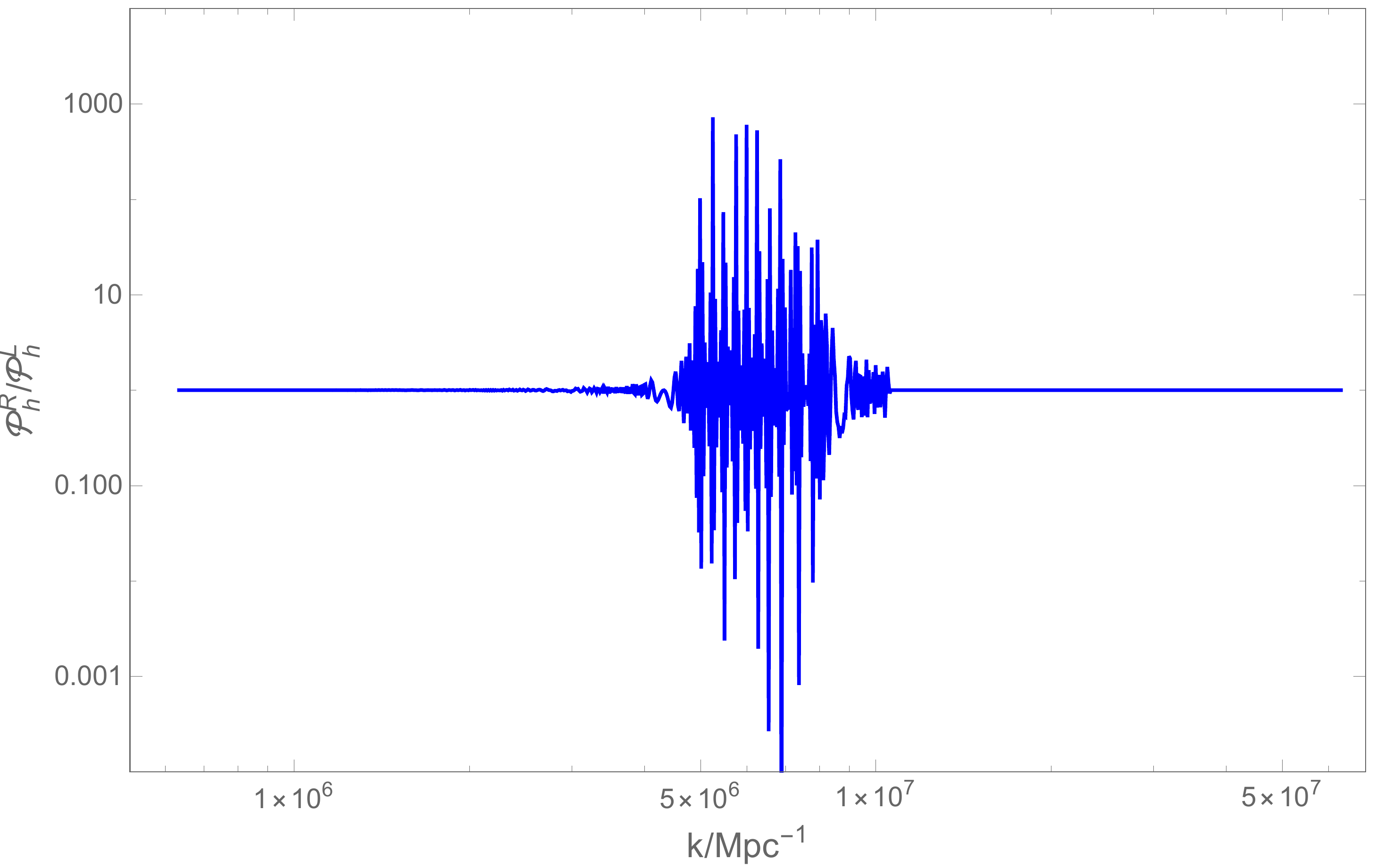}
\caption{\label{fig3} The ratio of $\mathcal{P}_h^\mathrm{R}$ to $\mathcal{P}_h^\mathrm{L}$ for case 1.}
\end{figure}

\begin{figure}
	\centering
	\includegraphics[width=0.8\textwidth ]{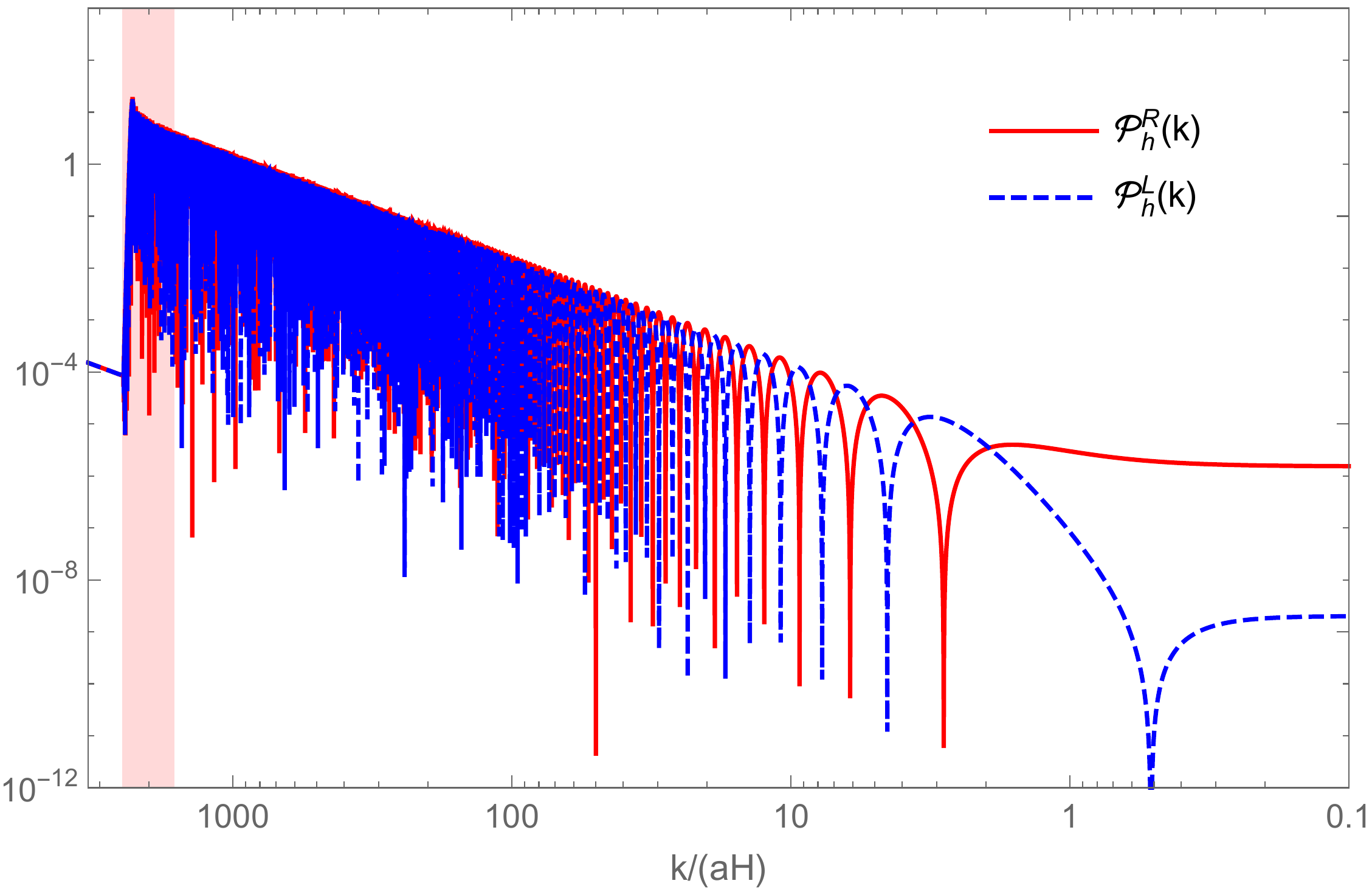}
	\caption{\label{fig4} The evolutions of $\mathcal{P}_h^\mathrm{R}(k)$ and $\mathcal{P}_h^\mathrm{L}(k)$ for a resonant $k$ mode as a function of $k/(aH)$ in case 1. The light-red region represents the period when the Chern-Simons term works.}
\end{figure}

Figure \ref{fig3} shows the ratio of power spectrum of the right-handed GWs $\mathcal{P}_h^\mathrm{R}=k^3|h_R(k)|^2/(2\pi^2)$ to that of the left-handed GWs $\mathcal{P}_h^\mathrm{L}=k^3|h_L(k)|^2/(2\pi^2)$ for case 1. From this figure, one can find that the power spectra of right-handed and left-handed GWs have markedly different amplitudes for the resonant modes, and the ratio rapidly oscillates with $k$. This phenomenon results from the significant difference of the amplitudes of two resonant polarization modes after they cross the horizon, which can be seen in Fig. \ref{fig4}. Apparently, the magnifications for both resonant polarization modes are almost identical when they are in horizon. The amplitude difference after horizon crossing between two resonant polarization modes originates mainly from the phase difference of $h_R$ and $h_L$ before the horizon
crossing (see Fig. \ref{fig4}), which results from the difference of their equations of motion \eqref{GW_EOM_h} when the Chern-Simons coupling is considered. Similar results can also be found in the other two cases. This provides an opportunity to directly detect the parity violation of the gravitational interaction in the future GW experiments.

\section{conclusions}
\label{sec5}
We have investigated the amplification of the parity-violating tensor perturbations during axion inflation with the gravitational Chern-Simons coupling. The Chern-Simons term is coupled to a periodic function $\vartheta(\phi)$ of the inflaton, so that the EOM of tensor perturbations can be transformed into the Mathieu equation. Tensor perturbations with modes in the narrow resonance band are exponentially amplified, which results in a narrow peak or a board plateau in $\Omega_{\mathrm{GW,0}}$, depending on the duration length of $\vartheta(\phi)$. The cutoff of $\Omega_{\mathrm{GW,0}}$ in the infrared region is a characteristic distinguishable from the stochastic GW background sourced by matter fields in Einstein gravity. Since the frequency of GWs depends on the comoving length scale at reentry, the amplified GWs can be detected by detectors sensitive to different frequencies. Moreover, the detection of this parity-violating GW signal may provide possible evidence of the Chern-Simons gravity with a periodic coupling function.

In the first Advanced LIGO observational run (aLIGO O1), the sum of all known noise sources cannot explain the measured sensitivity curve noise, especially below $100$Hz~\cite{Martynov:2016fzi}. This discrepancy has been reduced significantly in the aLIGO O2~\cite{Abbott:2019ebz}, but   some noise peaks remain unknown, for example the peak around $300$Hz observed by Livingston, Hanford and Virgo detectors. Interestingly, this noise peak could be explained by the resonance peak induced by the Chern-Simons term. This possibility should be examined by the next Advanced LIGO observational run with more detailed noise analysis.

\begin{acknowledgments}
We appreciate very much the insightful comments and helpful suggestions by anonymous referee, and  thank Xingyu Yang and Chang Liu for fruitful discussions.
C.F., P.W. and H.Y. were supported  by the National Key Research and Development Program of China Grant No. 2020YFC2201502, by the National Natural Science Foundation of China under Grants No. 11775077, No. 11435006, No. 11690034, and No. 11805063, and by the Science and Technology Innovation Plan of Hunan province under Grant No. 2017XK2019.
T.Z. was supported in part by National Natural Science Foundation of China under Grant No. 11675143, the Zhejiang Provincial Natural Science Foundation of China under Grants No. LR21A050001 and No. LY20A050002, and the Fundamental Research Funds for the Provincial Universities of Zhejiang in China with Grant No. RF-A2019015.

\end{acknowledgments}

\end{document}